\documentclass[aps,pra,twocolumn,superscriptaddress,nofootinbib,10pt]{revtex4-2}

\usepackage[T1]{fontenc}
\usepackage[english]{babel}

\usepackage{amsmath,amssymb}
\usepackage{graphicx}

\usepackage[
colorlinks=true,
breaklinks=true,
citecolor=blue,
linkcolor=blue,
urlcolor=blue
]{hyperref}

\newcommand{\rmd}{\mathrm{d}}
\newcommand{\rmi}{\mathrm{i}}
\newcommand{\rme}{\mathrm{e}}

\newcommand{\diag}{\operatorname{diag}}
\newcommand{\ad}{\operatorname{ad}}

\begin{document}
	
	\title{Improved convergence radius of the Fer expansion for Hermitian generators}
	
	\author{Lorenzo Bagnasacco}
	\email{lorenzo.bagnasacco@sns.it}
	\affiliation{NEST and Scuola Normale Superiore, Piazza dei Cavalieri 7, I-56126 Pisa, Italy}
	
	\author{Vittorio Giovannetti}
	\affiliation{NEST, Scuola Normale Superiore and Istituto Nanoscienze-CNR, Piazza dei Cavalieri 7, I-56126 Pisa, Italy}
	
	\begin{abstract}
		The Dyson series expands the propagator of a time-dependent Hamiltonian in powers of the Hamiltonian, but its truncations are in general not unitary. The Fer expansion writes the same propagator as an infinite product of matrix exponentials, each of them unitary, and its remainder decays doubly exponentially with the number of factors. Convergence, however, is guaranteed only within a finite radius: the time integral of the norm of the Hamiltonian must be smaller than $2$. This is the best value known to date. Here we improve it by about $30\%$, raising it to about $2.6058$. The result also holds for non-Hermitian Hamiltonians that are Hermitian with respect to a fixed metric.
	\end{abstract}
	
	\maketitle
	
	\begin{figure}[t]
		\centering
		\includegraphics[width=\linewidth]{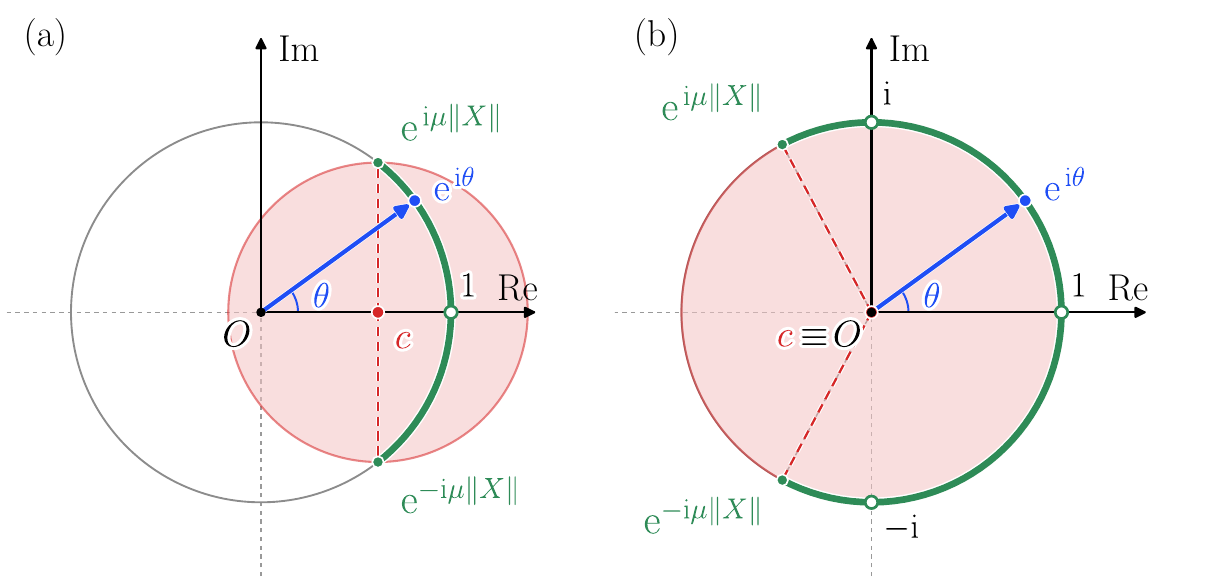}
		\caption{The spectrum of $\rme^{-\mu X}$ for anti-Hermitian $X$. The eigenvalues $\rme^{\rmi\theta}$ (blue) lie on the arc $|\theta|\le\mu\|X\|$ of the unit circle (green). The pink disc is the smallest disc centered at $c$ containing that arc; its radius, drawn as the red dashed segments from $c$ to the end points of the arc, is the quantity minimized over $c$ in Eq.~\eqref{eq:minmax}. (a) For $\mu\|X\|<\pi/2$ the optimal center is $c=\cos(\mu\|X\|)$, where the chord through the end points meets the real axis, and the radius is $\sin(\mu\|X\|)$, half that chord. (b) For $\mu\|X\|>\pi/2$ the arc contains the antipodal points $\pm\rmi$, so no disc of radius smaller than $1$ can contain it: the optimal center is the origin, $c\equiv O$, and the radius is $1$.}
		\label{fig:arc}
	\end{figure}
	
	\section{Introduction}
	\label{sec:intro}
	
	Time-dependent Hamiltonians are ubiquitous in quantum physics, and simulating the dynamics they generate is a central task for quantum algorithms~\cite{kieferova2019,berry2020,casares2024,fang2025}. The dynamics of a quantum system governed by a time-dependent Hamiltonian $H(t)$ is formally solved by the time-ordered exponential \mbox{$U(t)=\mathcal{T}\exp[-\rmi\int_0^t\rmd s\,H(s)]$}, whose expansion in powers of $H$ is the Dyson series~\cite{dyson}. This is only a formal solution: truncating the series destroys unitarity at every finite order. Reorganizing it into a form that keeps unitarity intact order by order is an old problem with two well-known answers.
	
	The first answer resums the Dyson series into a single exponential, $U(t)=\rme^{\Omega(t)}$, with $\Omega(t)$ an explicit anti-Hermitian operator built from nested commutators of $H$ at different times. This is the Magnus expansion~\cite{magnus,klarsfeld1989,blanes2009}: every truncation is unitary by construction, although $\Omega(t)$ is in general no easier to evaluate in closed form than the original problem. It also comes with a restriction on the time interval. The Magnus series is guaranteed to converge when $\int_0^t\|H(s)\|\,\rmd s<\pi$~\cite{moan2006,casas2007}, where $\|\cdot\|$ denotes the operator norm, and the constant $\pi$ cannot be replaced by a larger one: explicit generators exist for which the series diverges beyond it~\cite{moan2006}.
	
	The second answer keeps the propagator factorized into an ordered product of exponentials, each of them unitary, instead of resumming it into a single one. The standard constructions of this type are perturbative in the Hamiltonian, so that the $j$-th factor is of order $j$: the Trotter--Suzuki formulas that underlie much of digital quantum simulation~\cite{suzuki,lloyd1996,childs2021}, the Zassenhaus expansion~\cite{casas-zassenhaus}, and its continuous analogue due to Wilcox~\cite{wilcox,arnal2021}. Every truncation is again unitary, but the accuracy improves by one order at a time.
	
	The Fer expansion~\cite{fer,blanes1998} sits between these two answers. Like Magnus, it is built out of commutators of $H$ alone; like the product formulas, it represents the evolution operator as an ordered product rather than as a single exponential. It is not, however, organized in powers of the Hamiltonian. It peels off one exponential at a time, each generated by an effective Hamiltonian built recursively from the previous one: every truncation is manifestly unitary, and what is left over at each step is not an uncontrolled error but a genuine Schr\"odinger equation. What makes it attractive is that this leftover is quadratically small in the generator of the previous step: every new factor roughly doubles the order in time to which the truncated product reproduces the propagator, instead of adding one order to it~\cite{iserles1984,iserles2000}. The expansion was rediscovered in numerical analysis by Iserles~\cite{iserles1984} and later recognized as a Lie-group method~\cite{iserles2000,zanna1999}. It is used to build effective Hamiltonians in solid-state nuclear magnetic resonance~\cite{madhu2006,mananga2016}, in the same spirit as average-Hamiltonian theory~\cite{haeberlen1968,mananga2011} and dynamical decoupling~\cite{viola1999}, and to construct geometric integrators for differential equations on Lie groups~\cite{casas1996,zanna1999,zanna2001}.
	
	This fast convergence, however, is guaranteed only inside a finite region, and the region known so far is small. For a general generator $A(t)$, it was proved in Ref.~\cite{blanes1998} that the product converges whenever $\int_0^t\|A(s)\|\,\rmd s$ is smaller than about $0.8604$; the same work observes that when $A$ is anti-Hermitian, that is for a Hermitian Hamiltonian, every factor of the product is unitary and their argument then gives the larger value $2$. Neither constant appears to have been improved since, and the recent literature still quotes them~\cite{arnal2021}. Both are smaller than the radius $\pi$ of the Magnus expansion.
	
	In this work we improve the Hermitian value by about $30\%$. The whole gain comes from a freedom that the standard argument leaves unused. That argument measures the distance of a unitary from the identity; but any multiple of the identity would do just as well, since all of them commute with everything, and one is free to choose the one closest to the spectrum. Optimizing over that choice is an elementary problem of plane geometry, and it is what enlarges the radius. The gain relies on the unitarity of the factors: it extends to non-Hermitian Hamiltonians that are Hermitian with respect to a fixed metric, but not to general non-Hermitian ones, for which the classical one-step estimate of Ref.~\cite{blanes1998} is already optimal.
	
	The paper is organized as follows. Section~\ref{sec:fer} recalls the Fer recursion and fixes the notation. Section~\ref{sec:method} contains the argument: the one-step map (Sec.~\ref{sec:onestepmap}), the finite-angle estimate (Sec.~\ref{sec:angle}), the one-step bound it gives (Sec.~\ref{sec:onestep}), the fixed-point analysis that turns it into a radius (Sec.~\ref{sec:radius}), the convergence proof with its error bound (Sec.~\ref{sec:convergence}), and the comparison with the known value (Sec.~\ref{sec:comparison}). Section~\ref{sec:extensions} extends the result: Sec.~\ref{sec:shift} replaces the norm of the Hamiltonian with the half-width of its spectrum, Sec.~\ref{sec:metric} treats Hamiltonians that are Hermitian with respect to a fixed metric, and Sec.~\ref{sec:general} discusses general non-Hermitian Hamiltonians. Section~\ref{sec:conclusions} concludes.
	
	\section{The Fer expansion}
	\label{sec:fer}
	
	We briefly recall the construction of the Fer
	expansion, fixing notation and the recursive structure that will be used throughout.
	
	Given a time-dependent Hamiltonian $H(t)=H^{\dagger}(t)$, the time-evolution operator $U(t)$
	satisfies the Schr\"odinger equation ($\hbar=1$)
	\begin{equation}
		\frac{\rmd}{\rmd t}\,U(t) = -\rmi H(t)\,U(t)\;, \qquad U(0)=\openone\;.
		\label{eq:schrodinger}
	\end{equation}
	Following Refs.~\cite{fer,blanes1998}, an iterative solution can be constructed as
	an ordered product of matrix exponentials,
	\begin{equation}\label{eq:product}
		U(t) = \rme^{F_1(t)}\rme^{F_2(t)}\cdots\rme^{F_j(t)}\,U_{j+1}(t)
	\end{equation}
	where at every step the remainder $U_{j+1}(t)$ obeys an equation of the same
	Schr\"odinger form,
	\begin{equation}
		\frac{\rmd}{\rmd t}\,U_{j+1}(t) = -\rmi H_{j+1}(t)\,U_{j+1}(t)\;,
	\end{equation}
	governed by an effective Hamiltonian $H_{j+1}(t)$ generated recursively from
	the previous step through
	\begin{equation}
		\begin{split}
			H_{j+1}(t) &= \rme^{-F_j(t)}\,H_j(t)\,\rme^{F_j(t)}\\
			&\quad - \int_0^1\!\rmd \lambda\;\rme^{-\lambda F_j(t)}\,H_j(t)\,\rme^{\lambda F_j(t)}\;,\\
			F_j(t) &= -\rmi\int_0^t\!\rmd s\,H_j(s)\;,
		\end{split}
		\label{eq:fer-recursion}
	\end{equation}
	with the identification $H_1(t)\equiv H(t)$. This recursion is the
	defining feature of the Fer expansion: at each order $j$ the integral part of the dynamics is peeled off into $\rme^{F_j(t)}$, leaving an
	increasingly small remainder $H_{j+1}(t)$ to be treated at the next step~\cite{fer}. The recursion follows from the formula for the derivative of a matrix exponential~\cite{wilcox}.
	
	\begin{figure}[t]
		\centering
		\includegraphics[width=0.9\linewidth]{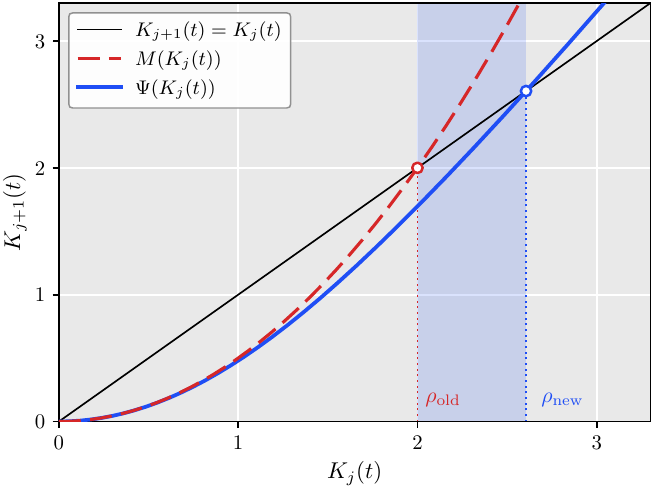}
		\caption{The recursion~\eqref{eq:Krec} seen as a map $K_j(t)\mapsto K_{j+1}(t)$: the finite-angle map $\Psi$ of Eq.~\eqref{eq:Psidef} (blue, solid) and the map $M$ of Eq.~\eqref{eq:Mdef} (red, dashed), which underlies the radius $2$ of Ref.~\cite{blanes1998}, together with the diagonal (thin, solid). The sufficient convergence radius is the positive fixed point, where the graph crosses the diagonal (circles): $\rho_{\mathrm{old}}=2$ for $M$ and $\rho_{\mathrm{new}}$, about $2.6058$, for $\Psi$, Eq.~\eqref{eq:rho}. Since $\Psi<M$, the crossing moves to the right: the shaded interval is the set of initial values $K_1(t)$ covered by the new bound but not the old one. Near the origin the two maps agree to leading order, so the rate of convergence is unaffected.}
		\label{fig:rho}
	\end{figure}
	
	\section{The method}
	\label{sec:method}
	
	Our goal is a sufficient condition on $H(t)$ under which the product~\eqref{eq:product} converges to $U(t)$. The strategy is the classical one of Ref.~\cite{blanes1998}: we bound the size of each effective Hamiltonian in terms of the previous one, reduce the recursion~\eqref{eq:fer-recursion} to a scalar map, and read off the convergence radius as the first non-trivial fixed point of that map.
	
	\subsection{The one-step map}
	\label{sec:onestepmap}
	
	The recursion~\eqref{eq:fer-recursion} has a simple structure: the first term is the Hamiltonian $H_j(t)$ seen in the frame rotated by $\rme^{F_j(t)}$; the second is what the exponential $\rme^{F_j(t)}$ has already accounted for, namely the average of the rotated Hamiltonian over the rotation. It is convenient to regard this step as the action of a linear map on the current effective Hamiltonian,
	\begin{equation}
		\begin{split}
			H_{j+1}(t) &= \mathcal{R}_{F_j(t)}\big(H_j(t)\big)\;,\\
			\mathcal{R}_X(Y) &\equiv \rme^{-X}\,Y\,\rme^{X} - \int_0^1\!\rmd \lambda\;\rme^{-\lambda X}\,Y\,\rme^{\lambda X}\;,
		\end{split}
		\label{eq:Tmap}
	\end{equation}
	where $X$ and $Y$ are auxiliary matrices. The whole expansion is then generated by the single map $\mathcal{R}_X$, applied at each order and each time $t$ with $X=F_j(t)$ and $Y=H_j(t)$. Two properties of $\mathcal{R}_X$ are all we need. First, it is linear in $Y$, and it vanishes when $Y$ commutes with $X$: this is why $H_{j+1}(t)$ is small whenever $H_j(t)$ nearly commutes with its own time integral $F_j(t)$, and it is the mechanism behind the quadratic convergence of the expansion. Second, since $H_j(t)$ is Hermitian, $F_j(t)=-\rmi\int_0^t\rmd s\,H_j(s)$ is anti-Hermitian, so that every $\rme^{\lambda F_j(t)}$ is unitary; by induction, $H_{j+1}(t)=\mathcal{R}_{F_j(t)}\big(H_j(t)\big)$ is Hermitian again.
	
	\subsection{A finite-angle estimate for unitary conjugation}
	\label{sec:angle}
	
	We now bound $\|\mathcal{R}_X(Y)\|$ in terms of $\|X\|$ and $\|Y\|$, for anti-Hermitian $X$. Writing the first term of Eq.~\eqref{eq:Tmap} as $\int_0^1\rmd\lambda\,\rme^{-X}Y\rme^{X}$, the map becomes the integral of the difference of two conjugations,
	\begin{equation}
		\mathcal{R}_X(Y) = \int_0^1\!\rmd\lambda\;\Big[\rme^{-X}\,Y\,\rme^{X} - \rme^{-\lambda X}\,Y\,\rme^{\lambda X}\Big]\;.
		\label{eq:Rdiff}
	\end{equation}
	The two conjugations share a common part. Set $\mu\equiv1-\lambda$. Exponentials of multiples of the same matrix commute, so $\rme^{-X}=\rme^{-\lambda X}\rme^{-\mu X}$ and $\rme^{X}=\rme^{\mu X}\rme^{\lambda X}$, and therefore
	\begin{equation}
		\rme^{-X}\,Y\,\rme^{X} - \rme^{-\lambda X}\,Y\,\rme^{\lambda X}
		= \rme^{-\lambda X}\Big[\rme^{-\mu X}\,Y\,\rme^{\mu X} - Y\Big]\rme^{\lambda X}\;.
		\label{eq:factor}
	\end{equation}
	The outer conjugation by the unitary $\rme^{\lambda X}$ does not change the operator norm. Taking the norm inside the integral in Eq.~\eqref{eq:Rdiff}, and integrating over $\mu$ instead of $\lambda$, we obtain
	\begin{equation}
		\big\|\mathcal{R}_X(Y)\big\| \le \int_0^1\!\rmd\mu\;\big\|\rme^{-\mu X}\,Y\,\rme^{\mu X} - Y\big\|\;.
		\label{eq:Rbound}
	\end{equation}
	The problem is thus reduced to estimating how far a single unitary conjugation moves $Y$, as a function of the rotation angle $\mu\|X\|$.
	
	Factoring out the unitary on the right,
	\begin{equation}
		\rme^{-\mu X}\,Y\,\rme^{\mu X} - Y = \big[\rme^{-\mu X},Y\big]\,\rme^{\mu X}\;,
		\label{eq:commutator}
	\end{equation}
	hence $\|\rme^{-\mu X}\,Y\,\rme^{\mu X} - Y\| \le \|[\rme^{-\mu X},Y]\|$.
	The elementary way to bound this commutator is to subtract the identity, $[\rme^{-\mu X},Y]=[\rme^{-\mu X}-\openone,Y]$, and use $\|\rme^{-\mu X}-\openone\|\le\mu\|X\|$: this gives $2\mu\|X\|\,\|Y\|$, a first-order estimate that grows without limit with the angle, and is the one behind the radius $2$ of Ref.~\cite{blanes1998}. The key observation is that the identity can be replaced by any multiple of it. A multiple of the identity commutes with everything, so that for any scalar $c$
	\begin{equation}
		\begin{split}
			\big[\rme^{-\mu X},Y\big] &= \big[\rme^{-\mu X}-c\openone,\,Y\big]\\
			&= \big(\rme^{-\mu X}-c\openone\big)\,Y - Y\,\big(\rme^{-\mu X}-c\openone\big)\;,
		\end{split}
		\label{eq:shiftid}
	\end{equation}
	and the triangle inequality, together with the submultiplicativity of the norm, gives
	\begin{equation}
		\big\|\big[\rme^{-\mu X},Y\big]\big\| \le 2\,\big\|\rme^{-\mu X}-c\openone\big\|\,\|Y\|
		\quad\text{for every } c\in\mathbb{C}\;.
		\label{eq:shift}
	\end{equation}
	The freedom in $c$ is where the gain comes from, and the best choice is dictated by the spectrum of $\rme^{-\mu X}$. Since $X$ is anti-Hermitian, it is diagonalized by a unitary $W$ and its eigenvalues are purely imaginary, $X=W\diag(\rmi\xi_1,\dots,\rmi\xi_d)W^\dagger$ with $\xi_k$ real and $|\xi_k|\le\|X\|$. In the same basis,
	\begin{equation}
		\begin{split}
			\rme^{-\mu X}-c\openone &= W\diag\big(\rme^{\rmi\theta_1}-c,\dots,\rme^{\rmi\theta_d}-c\big)W^\dagger\;,\\
			\theta_k &= -\mu\xi_k\;,\qquad |\theta_k|\le\mu\|X\|\;:
		\end{split}
		\label{eq:diagonal}
	\end{equation}
	the eigenvalues of $\rme^{-\mu X}$ lie on the arc $|\theta|\le\mu\|X\|$ of the unit circle (Fig.~\ref{fig:arc}). The unitary conjugation does not change the norm, and the norm of a diagonal matrix is the largest modulus of its entries. Hence
	\begin{equation}
		\big\|\rme^{-\mu X}-c\openone\big\| = \max_k\big|\rme^{\rmi\theta_k}-c\big| \le \max_{|\theta|\le\mu\|X\|}\big|\rme^{\rmi\theta}-c\big|\;,
		\label{eq:arc}
	\end{equation}
	where in the last step the maximum over the actual eigenvalues has been replaced by the maximum over the whole arc, so that the bound depends on $X$ only through $\|X\|$. Since $|\rme^{\rmi\theta}-c|$ is the Euclidean distance between the points $\rme^{\rmi\theta}$ and $c$ in the complex plane, the right-hand side is the radius of the smallest disc centered at $c$ that contains the arc. Equation~\eqref{eq:shift} holds for every $c$, so we are free to pick the best one:
	\begin{equation}
		\big\|\big[\rme^{-\mu X},Y\big]\big\| \le 2\,\|Y\|\;\min_{c\in\mathbb{C}}\;\max_{|\theta|\le\mu\|X\|}\big|\rme^{\rmi\theta}-c\big|\;.
		\label{eq:minmax}
	\end{equation}
	The minimum is the radius of the smallest disc containing the arc, an elementary problem of plane geometry with two cases (Fig.~\ref{fig:arc}). If $\mu\|X\|\le\pi/2$, the end points $\rme^{\pm\rmi\mu\|X\|}$ of the arc are $2\sin(\mu\|X\|)$ apart, so no disc containing them has radius smaller than $\sin(\mu\|X\|)$; on the other hand the disc of radius $\sin(\mu\|X\|)$ centered at $c=\cos(\mu\|X\|)$, the midpoint of the chord, contains the whole arc, because for $|\theta|\le\mu\|X\|$
	\begin{equation}
		\begin{split}
			\big|\rme^{\rmi\theta}-\cos(\mu\|X\|)\big|^2 &= 1+\cos^2(\mu\|X\|)\\
			&\quad-2\cos(\theta)\,\cos(\mu\|X\|)\\
			&\le 1-\cos^2(\mu\|X\|) = \sin^2(\mu\|X\|)\;,
		\end{split}
		\label{eq:disc}
	\end{equation}
	where we used $\cos(\theta)\ge\cos(\mu\|X\|)\ge0$, both inequalities holding because $|\theta|\le\mu\|X\|\le\pi/2$. The left-hand side is the squared distance between the point $\rme^{\rmi\theta}$ of the arc and $c$; it is largest at the end points, $\theta=\pm\mu\|X\|$, where it equals $\sin^2(\mu\|X\|)$. Taking the square root, every point of the arc is within distance $\sin(\mu\|X\|)$ of $c$, i.e. the arc lies inside the disc. The minimum is therefore $\sin(\mu\|X\|)$, attained at $c=\cos(\mu\|X\|)$. If instead $\mu\|X\|\ge\pi/2$, the arc contains the antipodal points $\pm\rmi$, which are $2$ apart, so no disc containing the arc has radius smaller than $1$; the unit disc, $c=0$, attains it. Altogether the minimum in Eq.~\eqref{eq:minmax} is $\sin(\min\{\mu\|X\|,\pi/2\})$, and Eq.~\eqref{eq:commutator} becomes
	\begin{equation}
		\big\|\rme^{-\mu X}\,Y\,\rme^{\mu X} - Y\big\| \le 2\sin\big(\min\{\mu\|X\|,\tfrac{\pi}{2}\}\big)\,\|Y\|\;.
		\label{eq:angle}
	\end{equation}
	The elementary estimate corresponds to the choice $c=1$ in Eq.~\eqref{eq:minmax}, which measures the arc by its chord $2\sin(\mu\|X\|/2)$, further bounded by its length $\mu\|X\|$. All three agree to first order in the angle: the improvement is invisible in the infinitesimal commutator bound and appears only at finite rotation angles.

	The estimate~\eqref{eq:angle} cannot be improved as long as only $\|X\|$ and $\|Y\|$ are used. By Eq.~\eqref{eq:commutator}, the left-hand side equals $\|[\rme^{-\mu X},Y]\|$, the norm of the derivation $Y\mapsto[\rme^{-\mu X},Y]$ applied to $Y$. Stampfli showed that the norm of this derivation, that is its maximum over all $Y$ with $\|Y\|=1$, is exactly $2\min_{c\in\mathbb{C}}\|\rme^{-\mu X}-c\openone\|$~\cite{stampfli1970}, so that Eq.~\eqref{eq:shift}, optimized over $c$, is attained by some $Y$. When $X$ has the eigenvalues $\pm\rmi\|X\|$, the two end points of the arc belong to the spectrum, the inequality in Eq.~\eqref{eq:arc} becomes an equality, and the right-hand side of Eq.~\eqref{eq:angle} is reached. The same quantity has a familiar meaning in quantum information. For a unitary $V$, the diamond distance between the channel $\sigma\mapsto V\sigma V^\dagger$ and the identity channel is $2\sqrt{1-\nu^2}$, with $\nu$ the distance from the origin of the numerical range of $V$~\cite{shatten-norm}. For $V=\rme^{-\mu X}$ the numerical range is the convex hull of the eigenvalues, and for $\mu\|X\|\le\pi/2$ its distance from the origin is at least $\cos(\mu\|X\|)$, the distance of the chord in Fig.~\ref{fig:arc}, panel (a): the diamond distance is then at most $2\sin(\mu\|X\|)$, with equality in the extremal case above, while for larger angles it can reach its maximum value $2$. The right-hand side of Eq.~\eqref{eq:angle} is therefore the largest diamond distance from the identity channel of a unitary channel generated by $X$, and the finite-angle estimate states that a unitary conjugation cannot move an operator, in operator norm, by more than this distance times $\|Y\|$.
	
	\subsection{The one-step bound}
	\label{sec:onestep}
	
	Inserting Eq.~\eqref{eq:angle} into Eq.~\eqref{eq:Rbound} and carrying out the integral over $\mu$ we arrive at the one-step bound
	\begin{equation}
		\big\|\mathcal{R}_X(Y)\big\| \le \psi\big(\|X\|\big)\,\|Y\|\;,
		\label{eq:psi}
	\end{equation}
	where
	\begin{equation}
		\begin{split}
			\psi\big(\|X\|\big) &\equiv \int_0^1\!\rmd\mu\;2\sin\big(\min\{\mu\|X\|,\tfrac{\pi}{2}\}\big)\\
			&=
			\begin{cases}
				\dfrac{2\,(1-\cos(\|X\|))}{\|X\|}\;, & 0\le\|X\|\le\pi/2\;,\\[2ex]
				2-\dfrac{\pi-2}{\|X\|}\;, & \|X\|\ge\pi/2\;,
			\end{cases}
		\end{split}
		\label{eq:psidef}
	\end{equation}
	with $\psi(0)=0$. For $\|X\|\le\pi/2$ the integrand is $2\sin(\mu\|X\|)$ on the whole interval; for $\|X\|>\pi/2$ it is $2\sin(\mu\|X\|)$ up to $\mu=\pi/(2\|X\|)$ and equal to $2$ afterwards, which gives the second branch. Both branches equal $4/\pi$ at $\|X\|=\pi/2$, so $\psi$ is continuous; it is increasing, being the integral of an increasing function of $\|X\|$, and it tends to $2$ as $\|X\|\to\infty$. Most importantly,
	\begin{equation}
		\psi\big(\|X\|\big) < \|X\| \qquad\text{for every } \|X\|>0\;,
		\label{eq:psilta}
	\end{equation}
	because $2(1-\cos(\|X\|))<\|X\|^2$ for $0<\|X\|\le\pi/2$, and $\|X\|-\psi(\|X\|)=[(\|X\|-1)^2+\pi-3]/\|X\|>0$ for $\|X\|\ge\pi/2$. For small $\|X\|$ one has $\psi(\|X\|)=\|X\|-\|X\|^3/12+O(\|X\|^5)$: the finite-angle bound coincides with the elementary one at first order and improves on it at every finite $\|X\|$. Applied to the recursion~\eqref{eq:fer-recursion}, with $X=F_j(t)$ anti-Hermitian and $Y=H_j(t)$, Eq.~\eqref{eq:psi} gives at every time $t$
	\begin{equation}
		\big\|H_{j+1}(t)\big\| \le \psi\big(\|F_j(t)\|\big)\,\big\|H_j(t)\big\|\;,
		\label{eq:Hstep}
	\end{equation}
	which is the one-step bound on which the convergence radius will be built.
	
	\subsection{From the one-step bound to the convergence radius}
	\label{sec:radius}
	
	The bound~\eqref{eq:Hstep} controls the size of $H_{j+1}(t)$ at a given time through the size of $H_j(t)$ at the same time and through $\|F_j(t)\|$, which is itself an integral of $H_j$ over earlier times. The natural scalar quantity to follow through the recursion is therefore
	\begin{equation}
		K_j(t) \equiv \int_0^t\!\rmd s\;\big\|H_j(s)\big\|\;,
		\qquad
		K_1(t) = \int_0^t\!\rmd s\;\big\|H(s)\big\|\;,
		\label{eq:Kdef}
	\end{equation}
	a non-decreasing function of $t$ with $K_j(0)=0$ and $\dot K_j(t)=\|H_j(t)\|$. Since $F_j(t)=-\rmi\int_0^t\rmd s\,H_j(s)$, the triangle inequality gives $\|F_j(t)\|\le K_j(t)$, and because $\psi$ is increasing, Eq.~\eqref{eq:Hstep} implies
	\begin{equation}
		\dot K_{j+1}(t) = \big\|H_{j+1}(t)\big\| \le \psi\big(K_j(t)\big)\,\dot K_j(t)\;.
		\label{eq:Kdot}
	\end{equation}
	The right-hand side is a total derivative: $\psi(K_j(t))\,\dot K_j(t)=\frac{\rmd}{\rmd t}\Psi(K_j(t))$, with
	\begin{equation}
		\Psi(x) \equiv \int_0^x\!\rmd u\;\psi(u)\;.
		\label{eq:Psidef}
	\end{equation}
	Integrating Eq.~\eqref{eq:Kdot} from $0$ to $t$, where both sides vanish, we obtain the scalar recursion
	\begin{equation}
		K_{j+1}(t) \le \Psi\big(K_j(t)\big)\;.
		\label{eq:Krec}
	\end{equation}
	The function $\Psi$ is explicit. For $x\le\pi/2$ it is $\Psi(x)=2\int_0^x\rmd u\,(1-\cos(u))/u$, not elementary but easily evaluated, with $\Psi(\pi/2)$ equal to about $1.1136$. For $x\ge\pi/2$ the second branch of $\psi$ integrates to
	\begin{equation}
		\Psi(x) = \Psi\big(\tfrac{\pi}{2}\big) + 2\Big(x-\frac{\pi}{2}\Big) - (\pi-2)\ln\!\Big(\frac{2x}{\pi}\Big)\;.
		\label{eq:Psi2}
	\end{equation}
	
	The convergence of the Fer expansion is now a question about the iteration of $\Psi$. Its graph starts at $\Psi(0)=0$ with zero slope, $\Psi'(0)=\psi(0)=0$, and grows like $2x$ for large $x$, since $\psi\to2$. It is convex, because $\Psi'=\psi$ is increasing. Hence $\Psi(x)-x$ is a convex function that vanishes at $x=0$, starts decreasing with slope $-1$, and tends to $+\infty$: it has exactly one positive zero, which we call $\rho$,
	\begin{equation}
		\Psi(\rho) = \rho\;,
		\qquad
		\Psi(x) < x \quad\text{for } 0<x<\rho\;.
		\label{eq:fixedpoint}
	\end{equation}
	Since $\Psi(\pi/2)$, about $1.1136$, is smaller than $\pi/2$, the fixed point lies on the branch~\eqref{eq:Psi2}, where the equation $\Psi(\rho)=\rho$ becomes $\rho=\pi-\Psi(\pi/2)+(\pi-2)\ln(2\rho/\pi)$ and is solved numerically:
	\begin{equation}
		\rho \approx 2.6058\;.
		\label{eq:rho}
	\end{equation}
	
	\subsection{Convergence and error bound}
	\label{sec:convergence}
	
	Suppose now that $K_1(t)<\rho$. Then, by Eqs.~\eqref{eq:Krec} and~\eqref{eq:fixedpoint}, $K_2(t)\le\Psi(K_1(t))<K_1(t)$, and inductively $K_{j+1}(t)\le\Psi(K_j(t))<K_j(t)$ for every $j$: the sequence $K_j(t)$ is decreasing and non-negative, so it has a limit, and the limit $\ell$ satisfies $\ell\le\Psi(\ell)$ with $\ell<\rho$, which forces $\ell=0$. Thus $K_j(t)\to0$. The convergence is in fact very fast: by Eq.~\eqref{eq:psilta}, $\Psi(x)\le\int_0^x u\,\rmd u=x^2/2$, so once $K_j(t)$ drops below $1$ the recursion~\eqref{eq:Krec} gives $K_{j+1}(t)\le K_j(t)^2/2$, and $K_j(t)$ decays doubly exponentially in $j$. In particular $\sum_j K_j(t)<\infty$.
	
	This is what is needed for the product~\eqref{eq:product} to converge. Each factor satisfies $\|\rme^{F_j(t)}-\openone\|\le\|F_j(t)\|\le K_j(t)$, because $F_j(t)$ is anti-Hermitian, and a product of unitaries $\prod_j\rme^{F_j(t)}$ with $\sum_j\|\rme^{F_j(t)}-\openone\|<\infty$ converges in operator norm. Its limit is the exact propagator: after $N$ factors,
	\begin{equation}
		U(t) = \rme^{F_1(t)}\rme^{F_2(t)}\cdots\rme^{F_N(t)}\,U_{N+1}(t)\;,
	\end{equation}
	where $U_{N+1}(t)$ is the propagator generated by $H_{N+1}(t)$. Integrating its Schr\"odinger equation,
	\begin{equation}
		U_{N+1}(t)-\openone=-\rmi\int_0^t\!\rmd s\,H_{N+1}(s)\,U_{N+1}(s)\;,
	\end{equation}
	and, since $U_{N+1}(s)$ is unitary, $\|U_{N+1}(t)-\openone\|\le K_{N+1}(t)$. Multiplying by the unitary product of the first $N$ factors does not change the norm, and we arrive at the explicit error bound
	\begin{equation}
		\Big\|U(t)-\rme^{F_1(t)}\cdots\rme^{F_N(t)}\Big\| \le K_{N+1}(t) \le \Psi^{\circ N}\big(K_1(t)\big)\;,
		\label{eq:error}
	\end{equation}
	with $\Psi^{\circ N}$ the $N$-fold iterate of $\Psi$. We have thus proved that the Fer expansion of a Hermitian Hamiltonian converges in operator norm whenever
	\begin{equation}
		\int_0^t\!\rmd s\;\big\|H(s)\big\| < \rho \approx 2.6058\;,
		\label{eq:main}
	\end{equation}
	with the truncation error controlled by Eq.~\eqref{eq:error}. Since a multiple of the identity added to $H(t)$ changes only the first factor of the product, and only by a global phase, $\|H(s)\|$ in Eq.~\eqref{eq:main} can be replaced by the half-width of the spectrum of $H(s)$, as shown in Sec.~\ref{sec:shift}.
	
	\subsection{Comparison with the known radius}
	\label{sec:comparison}
	
	With the elementary estimate $\psi(\|X\|)\to\|X\|$, the same construction gives the scalar map behind the radius $2$ of Ref.~\cite{blanes1998}. We denote it by $M$, the symbol used there for the recursion $K_{j+1}=M(K_j)$; for a Hermitian Hamiltonian it reads
	\begin{equation}
		M(x) = \int_0^x\!\rmd u\;u = \frac{x^2}{2}\;,
		\label{eq:Mdef}
	\end{equation}
	whose positive fixed point is $\rho_{\mathrm{old}}=2$. The finite-angle map lies strictly below, $\Psi(x)<M(x)$ for $x>0$ by Eq.~\eqref{eq:psilta}, so its graph meets the diagonal later, at $\rho_{\mathrm{new}}$, about $2.6058$, that is about $30\%$ further (Fig.~\ref{fig:rho}). Near $x=0$ the two maps agree to leading order, $\Psi(x)=M(x)-x^4/48+O(x^6)$, so the asymptotic doubly exponential rate is the same in both cases: what the finite-angle estimate changes is not the speed of convergence once it has set in, but the size of the initial data for which it is guaranteed to set in at all.
	
	\section{Extensions and limits}
	\label{sec:extensions}

	We now discuss how the result changes when the Hamiltonian contains a multiple of the identity, and what survives of it for non-Hermitian Hamiltonians.

	\subsection{Shifting the Hamiltonian by a multiple of the identity}
	\label{sec:shift}

	The condition~\eqref{eq:main} is written in terms of the operator norm $\|H(s)\|$. This quantity changes if a multiple of the identity is added to the Hamiltonian, although the dynamics changes only by a global phase: adding a large constant to $H(s)$ makes $\|H(s)\|$ large without changing anything physical. Removing such a term is the simplest instance of the preliminary transformations commonly used to improve the accuracy and the convergence domain of exponential expansions~\cite{blanes2009}. Here we show that the Fer expansion is unaffected by the shift, except for a phase in its first factor, and we use this to replace $\|H(s)\|$ in Eq.~\eqref{eq:main} with the half-width of the spectrum of $H(s)$.

	We subtract from $H(t)$ a multiple of the identity with a real, time-dependent coefficient $\gamma(t)$, and call $\Gamma(t)$ the phase it accumulates,
	\begin{equation}
		\widetilde H(t) = H(t)-\gamma(t)\,\openone\;,\qquad \Gamma(t)=\int_0^t\!\rmd s\;\gamma(s)\;.
		\label{eq:shiftH}
	\end{equation}
	The shifted Hamiltonian generates the same dynamics up to this phase. Since $\rme^{\rmi\Gamma(t)}$ is a scalar, differentiating $\rme^{\rmi\Gamma(t)}U(t)$ gives $\rmi\gamma(t)\,\rme^{\rmi\Gamma(t)}U(t)-\rmi H(t)\,\rme^{\rmi\Gamma(t)}U(t)=-\rmi\widetilde H(t)\,\rme^{\rmi\Gamma(t)}U(t)$, so that the propagator $\widetilde U(t)$ generated by $\widetilde H(t)$ is
	\begin{equation}
		\widetilde U(t)=\rme^{\rmi\Gamma(t)}\,U(t)\;.
		\label{eq:phase}
	\end{equation}

	We now compare the Fer expansions of $H$ and $\widetilde H$. The first generator of the shifted problem differs from $F_1(t)$ only by a multiple of the identity, $-\rmi\int_0^t\rmd s\,\widetilde H(s)=F_1(t)+\rmi\Gamma(t)\,\openone$. The second step uses the map $\mathcal{R}_X$ of Eq.~\eqref{eq:Tmap}, which is insensitive to multiples of the identity in both its arguments: for complex $\alpha$ and $\beta$,
	\begin{align}
		\mathcal{R}_{X+\rmi\alpha\openone}(Y) &= \mathcal{R}_X(Y)\;, \label{eq:invX}\\
		\mathcal{R}_X(Y+\beta\openone) &= \mathcal{R}_X(Y)\;. \label{eq:invY}
	\end{align}
	Using first Eq.~\eqref{eq:invX} and then Eq.~\eqref{eq:invY}, the effective Hamiltonian produced by the shifted problem at the second step is
	\begin{equation}
			H_2(t)=\mathcal{R}_{F_1(t)+\rmi\Gamma(t)\openone}\big(\widetilde H(t)\big)=\mathcal{R}_{F_1(t)}\big(H(t)\big)\;,
		\label{eq:sameH2}
	\end{equation}
	the same as for $H$. From here on the two recursions~\eqref{eq:fer-recursion} are identical, and they produce the same $H_j(t)$ and $F_j(t)$ for every $j\ge2$. The two Fer products therefore differ only in their first factor, and there only by the phase, $\rme^{F_1(t)+\rmi\Gamma(t)\openone}=\rme^{\rmi\Gamma(t)}\rme^{F_1(t)}$.

	This is what allows us to compare the truncation errors. After $N$ factors, the Fer approximation of $\widetilde U(t)$ is $\rme^{\rmi\Gamma(t)}\rme^{F_1(t)}\cdots\rme^{F_N(t)}$, that is the Fer approximation of $U(t)$ multiplied by the phase. Since by Eq.~\eqref{eq:phase} the exact propagator $\widetilde U(t)$ carries the same phase, the phase can be factored out of the difference,
	\begin{equation}
		\begin{split}
			&\widetilde U(t)-\rme^{\rmi\Gamma(t)}\rme^{F_1(t)}\cdots\rme^{F_N(t)}\\
			&\qquad=\rme^{\rmi\Gamma(t)}\Big[U(t)-\rme^{F_1(t)}\cdots\rme^{F_N(t)}\Big]\;,
		\end{split}
		\label{eq:sameerror}
	\end{equation}
	and, since $|\rme^{\rmi\Gamma(t)}|=1$, the two errors have the same norm for every $N$. The error of the ordinary Fer expansion of $H$ can then be bounded by applying the results of Sec.~\ref{sec:convergence} to the Hermitian Hamiltonian $\widetilde H$, which gives
	\begin{equation}
		\begin{split}
			&\Big\|U(t)-\rme^{F_1(t)}\cdots\rme^{F_N(t)}\Big\|\\
			&\qquad\le \Psi^{\circ N}\Big(\int_0^t\!\rmd s\;\big\|H(s)-\gamma(s)\openone\big\|\Big)\;.
		\end{split}
		\label{eq:errorshift}
	\end{equation}
	The Fer expansion of $H$ thus converges whenever $\int_0^t\rmd s\,\|H(s)-\gamma(s)\openone\|<\rho$. This holds for every real $\gamma(t)$, and the choice $\gamma=0$ gives back Eqs.~\eqref{eq:error} and~\eqref{eq:main}.

	It remains to choose $\gamma$. The best choice makes the integral in Eq.~\eqref{eq:errorshift} as small as possible, and since $\gamma(s)$ can be chosen independently at every time, it is enough to minimize $\|H(s)-\gamma(s)\openone\|$ at each $s$. The eigenvalues of $H(s)-\gamma\openone$ are those of $H(s)$ shifted by $-\gamma$, so that $\|H(s)-\gamma\openone\|=\max\{|\lambda_{\max}(s)-\gamma|,|\lambda_{\min}(s)-\gamma|\}$, with $\lambda_{\max}(s)$ and $\lambda_{\min}(s)$ the largest and smallest eigenvalues of $H(s)$. This is the distance from $\gamma$ to the farther end of the interval $[\lambda_{\min}(s),\lambda_{\max}(s)]$, and it is smallest when $\gamma$ is the midpoint, $\gamma(s)=[\lambda_{\max}(s)+\lambda_{\min}(s)]/2$, where it equals the half-width of the spectrum, $[\lambda_{\max}(s)-\lambda_{\min}(s)]/2$. This is the one-dimensional counterpart of the smallest disc of Sec.~\ref{sec:angle}: there the center $c$ was chosen for the eigenvalues of $\rme^{-\mu X}$ on the unit circle, here the center $\gamma$ is chosen for the eigenvalues of $H(s)$ on the real line. With this choice the convergence condition becomes
	\begin{equation}
		\int_0^t\!\rmd s\;\frac{\lambda_{\max}(s)-\lambda_{\min}(s)}{2} < \rho \approx 2.6058\;,
		\label{eq:mainshift}
	\end{equation}
	and the error is bounded by $\Psi^{\circ N}$ of the same integral. The radius $\rho$ is unchanged, since it is a property of the map $\Psi$; what changes is only the quantity that is integrated.

	The new condition is never more restrictive than the old one. Since $|\lambda_{\max}(s)|$ and $|\lambda_{\min}(s)|$ are both at most $\|H(s)\|$, the half-width never exceeds the norm, $[\lambda_{\max}(s)-\lambda_{\min}(s)]/2\le\|H(s)\|$, so every Hamiltonian that satisfies Eq.~\eqref{eq:main} also satisfies Eq.~\eqref{eq:mainshift}. The two conditions coincide when $\lambda_{\min}(s)=-\lambda_{\max}(s)$, that is when the spectrum is already centered at zero, and the gain is largest for Hamiltonians of definite sign. If $H(s)$ is positive semidefinite, then $\lambda_{\min}(s)\ge0$ and the half-width is at most $\|H(s)\|/2$, with equality when $H(s)$ has a zero eigenvalue. For such Hamiltonians Eq.~\eqref{eq:mainshift} holds whenever $\int_0^t\rmd s\,\|H(s)\|<2\rho$, that is below about $5.21$.

	The same shift can be applied to the Magnus expansion. All its terms beyond the first are built from commutators, which do not see multiples of the identity, so the Magnus exponents of $H$ and $\widetilde H$ differ only by the scalar $\rmi\Gamma(t)\openone$, and the condition of Refs.~\cite{moan2006,casas2007} also holds with the half-width in place of $\|H(s)\|$. The comparison between $\rho$ and the Magnus radius $\pi$ made in Sec.~\ref{sec:intro} is therefore unaffected. What is specific to the Fer expansion is that the shift, although it enters the first factor, leaves all the following ones exactly unchanged.

	\subsection{Hamiltonians that are Hermitian in a fixed metric}
	\label{sec:metric}

	The recursion~\eqref{eq:fer-recursion} does not use the Hermiticity of $H(t)$, and defines the Fer expansion of any time-dependent matrix. Hermiticity enters the proof of Sec.~\ref{sec:method} only through the unitarity of the exponentials $\rme^{\lambda F_j(t)}$, so the result extends to non-Hermitian Hamiltonians that become Hermitian after a fixed change of basis. These are the quasi-Hermitian Hamiltonians~\cite{scholtz1992,mostafazadeh2010}, which include $\mathcal{PT}$-symmetric Hamiltonians with unbroken symmetry~\cite{bender2007}: there is a positive matrix $P$, the metric, such that
	\begin{equation}
		H^\dagger(t)\,P=P\,H(t)\;,
		\label{eq:quasiH}
	\end{equation}
	and we require the same $P$ at all times. With $S=P^{1/2}$, Eq.~\eqref{eq:quasiH} states that $h(t)=S\,H(t)\,S^{-1}$ is Hermitian.

	The Fer expansion is covariant under this change of basis. Since $\rme^{SXS^{-1}}=S\,\rme^{X}S^{-1}$, the map~\eqref{eq:Tmap} satisfies $\mathcal{R}_{SXS^{-1}}(SYS^{-1})=S\,\mathcal{R}_X(Y)\,S^{-1}$, and the recursion started from $h$ produces $S H_j(t) S^{-1}$ and $S F_j(t) S^{-1}$ at every step. The propagator generated by $h(t)$ is the unitary $u(t)=S\,U(t)\,S^{-1}$, so that
	\begin{equation}
		\begin{split}
			&U(t)-\rme^{F_1(t)}\cdots\rme^{F_N(t)}\\
			&\qquad=S^{-1}\Big[u(t)-\rme^{SF_1(t)S^{-1}}\cdots\rme^{SF_N(t)S^{-1}}\Big]S\;.
		\end{split}
	\end{equation}
	The bracket is the error of the Fer expansion of the Hermitian Hamiltonian $h$, bounded in Sec.~\ref{sec:convergence}. Since $h(s)$ is Hermitian and has the same eigenvalues $\lambda_k(s)$ as $H(s)$, these are real and $\|h(s)\|=\max_k|\lambda_k(s)|$. Hence
	\begin{equation}
		\begin{split}
			&\Big\|U(t)-\rme^{F_1(t)}\cdots\rme^{F_N(t)}\Big\|\\
			&\qquad\le \|S\|\,\|S^{-1}\|\;\Psi^{\circ N}\Big(\int_0^t\!\rmd s\;\max_k|\lambda_k(s)|\Big)\;,
		\end{split}
	\end{equation}
	and the Fer expansion of $H$ converges whenever
	\begin{equation}
		\int_0^t\!\rmd s\;\max_k\big|\lambda_k(s)\big| < \rho \approx 2.6058\;.
		\label{eq:mainquasi}
	\end{equation}
	The metric enters only the prefactor $\|S\|\,\|S^{-1}\|$ of the error, not the convergence condition. Since the largest modulus of the eigenvalues never exceeds the norm, Eq.~\eqref{eq:mainquasi} holds in particular whenever $\int_0^t\rmd s\,\|H(s)\|<\rho$: the condition~\eqref{eq:main} extends unchanged to these Hamiltonians. The shift of Sec.~\ref{sec:shift}, with real $\gamma(t)$, preserves Eq.~\eqref{eq:quasiH} and replaces $\max_k|\lambda_k(s)|$ with the half-width of the spectrum. The metric must be time independent: a time-dependent $S(t)$ adds the term $\rmi\dot S(t)S^{-1}(t)$ to $h(t)$, which is not Hermitian in general, and the argument does not apply.

	\subsection{General non-Hermitian Hamiltonians}
	\label{sec:general}

	For a general non-Hermitian $H(t)$ both ingredients of Sec.~\ref{sec:angle} are lost: the conjugations by $\rme^{\lambda F_j(t)}$ are no longer isometries, and the distance of $\rme^{-\mu X}$ from the multiples of the identity is no longer fixed by its spectrum. The best one-step estimate is then the classical one~\cite{blanes1998}, as we now show. Expanding the exponentials in Eq.~\eqref{eq:Tmap} in powers of $\ad_X\equiv[X,\,\cdot\,]$,
	\begin{equation}
		\mathcal{R}_X=\sum_{n=1}^{\infty}\frac{(-1)^n\,n}{(n+1)!}\,\ad_X^{\,n}\;.
		\label{eq:Rseries}
	\end{equation}
	Since $\ad_X$ does not change when a multiple of the identity is subtracted from $X$, $\|\ad_X\|\le2\|X-c\openone\|$ for every complex $c$, and summing the series gives
	\begin{equation}
		\big\|\mathcal{R}_X(Y)\big\|\le\varphi\big(\delta(X)\big)\,\|Y\|\;,
		\label{eq:phibound}
	\end{equation}
	with $\varphi(x)=\rme^{2x}-(\rme^{2x}-1)/(2x)$ and $\delta(X)=\min_{c\in\mathbb{C}}\|X-c\openone\|$. With $c=0$, $\varphi$ is the coefficient used for general generators in Ref.~\cite{blanes1998}, whose integral $\int_0^x\rmd u\,\varphi(u)$ has its fixed point at about $0.8604$.

	The estimate~\eqref{eq:phibound} cannot be improved. For the $2\times2$ matrices
	\begin{equation}
		X=x\begin{pmatrix}1&0\\0&-1\end{pmatrix}\;,\qquad Y=\begin{pmatrix}0&0\\1&0\end{pmatrix}\;,
	\end{equation}
	one has $\rme^{-\lambda X}Y\rme^{\lambda X}=\rme^{2\lambda x}\,Y$, hence $\mathcal{R}_X(Y)=\varphi(x)\,Y$, with $\|Y\|=1$ and $\delta(X)=x$. This $X$ corresponds to a Hamiltonian with the purely imaginary eigenvalues $\pm\rmi x$, describing gain and loss. For anti-Hermitian $X$ the coefficient is instead $\psi(\|X\|)$ of Eq.~\eqref{eq:psidef}, which never exceeds $2$, while $\varphi$ grows exponentially: the finite-angle improvement relies on unitarity and has no counterpart for general Hamiltonians. The value of about $0.8604$ is therefore the best that one-step estimates based on norms alone can give. Whether the full recursion converges beyond it remains open, since in the example $X$ and $Y$ are chosen independently, while in the recursion $F_j(t)$ is the time integral of $H_j(t)$.

	The shift of Sec.~\ref{sec:shift} still applies, with a multiple $\zeta(t)\,\openone$ of the identity whose coefficient $\zeta(t)$, unlike $\gamma(t)$ in Sec.~\ref{sec:shift}, can be complex: there $\gamma(t)$ had to be real to keep the shifted Hamiltonian Hermitian, while here there is no Hermiticity to preserve. The factor $\exp[\rmi\int_0^t\rmd s\,\zeta(s)]$ that relates the two propagators and the two Fer products no longer has modulus one, but it is the same for both and does not affect convergence. The Fer expansion of a general $H(t)$ therefore converges whenever
	\begin{equation}
		\int_0^t\!\rmd s\;\big\|H(s)-\zeta(s)\,\openone\big\| \lesssim 0.8604
		\label{eq:maingeneral}
	\end{equation}
	for some complex $\zeta(t)$, and the best choice minimizes $\|H(s)-\zeta(s)\openone\|$ at each time.

	\section{Conclusions}
	\label{sec:conclusions}
	
	We have improved the sufficient convergence radius of the Fer expansion for Hermitian generators from the value $\rho_{\mathrm{old}}=2$ of Ref.~\cite{blanes1998} to $\rho_{\mathrm{new}}$, about $2.6058$, Eq.~\eqref{eq:main}. The improvement comes from a single geometric observation. The spectrum of a unitary conjugation $\rme^{-\mu X}$ lies on an arc of the unit circle, and the size of the commutator $[\rme^{-\mu X},Y]$ is controlled by the distance of that arc from any multiple $c\openone$ of the identity, not only from $\openone$ itself. Optimizing over the free parameter $c$ replaces the length of the arc by the radius of the smallest disc containing it, Eq.~\eqref{eq:angle}, and turns the elementary one-step coefficient $\|X\|$ into $\psi(\|X\|)<\|X\|$. Everything else is the standard scalar recursion, whose fixed point moves from $\rho_{\mathrm{old}}$ to $\rho_{\mathrm{new}}$.
	
	The gain is about $30\%$ in the radius, while the rate of convergence is unchanged: $\psi$ agrees with the elementary estimate to first order, so near the origin $\Psi$ still behaves as the map $M$ of Ref.~\cite{blanes1998} and the remainder still decays doubly exponentially. What it enlarges is the set of Hamiltonians for which that decay is guaranteed. The same radius holds for non-Hermitian Hamiltonians that are Hermitian with respect to a fixed metric, with the norm replaced by the largest modulus of the eigenvalues (Sec.~\ref{sec:metric}). For general non-Hermitian Hamiltonians, instead, the classical one-step estimate cannot be improved, so that the value of about $0.8604$ found in Ref.~\cite{blanes1998} is the best that such estimates can give (Sec.~\ref{sec:general}).

	The sharp convergence radius of the Fer expansion for Hermitian generators remains unknown. Since the one-step estimate~\eqref{eq:angle} is optimal~\cite{stampfli1970}, and coincides with the largest diamond distance between the corresponding unitary channel and the identity~\cite{shatten-norm}, a further improvement cannot come from better norm bounds on single conjugations: it requires keeping track of the structure of the effective Hamiltonians across the recursion, for instance of the correlations between $H_j(t)$ and its time integral $F_j(t)$.
	
	\begin{acknowledgments}
	Large language models, ChatGPT SOL (OpenAI) and Claude Code (Anthropic), were used as research tools in this work. An AI system suggested a technical step that sharpens the one-step estimate at the core of the convergence proof; it contributed a technique, not the solution of the problem. The technique, which measures the distance of a unitary from the multiples of the identity, was already known in the theory of quantum channels; we transferred it to the Fer expansion and adapted it to the convergence analysis. The formulation of the problem, the interpretation of this step as a geometric statement about the spectrum of unitary operators, and the convergence analysis built on it are due to the authors, within a research program on the Fer expansion and related product expansions that both authors have pursued for several years. Language models were also used to edit the manuscript. All results were checked by hand by the authors, and some were derived without AI assistance. The authors take full responsibility for the content.
	\end{acknowledgments}

	\bibliographystyle{prsty-title-hyperref}
	\bibliography{bib}
	
\end{document}